\documentclass[apj]{emulateapj}
\usepackage{amsmath}
\usepackage{apjfonts}
\usepackage{multirow}
\usepackage{array}

\newcommand{\DDir}{\relax{D\kern-.7em{/}}}

\newcommand{\inv}[1]{\frac{1}{#1}}

\newcommand{\ra}{\rightarrow}

\newcommand{\xra}{\xrightarrow}

\newcommand{\be}{\begin{equation}}
\newcommand{\ee}{\end{equation}}
\newcommand{\bea}{\begin{equation*}}
\newcommand{\eea}{\end{equation*}}

\newcommand{\abs}[1]{\left\vert#1\right\vert}

\newcommand{\pr}{\partial}

\newcommand{\nin}{\relax{\in\kern-.8em{/}}}

\newcommand{\bt}{\beta}

\newcommand{\lm}{\lambda}

\newcommand{\de}{\delta}

\newcommand{\vt}{\textrm{v}}

\newcommand{\cm}{\mbox{ cm}}

\newcommand{\gr}{\mbox{g}}

\newcommand{\sref}{\S~\ref}

\newcommand{\SWA}{SW ansatz}
\newcommand{\msheff}{m_{\rm sh, eff}}
\newcommand{\vsheff}{\vt_{\rm sh,eff}}

\newcommand{\tref}{t_{\rm ref}}
\newcommand{\s}{\rm sh}

\begin{document}
\title{Non-relativistic radiation mediated shock breakouts: \\ I. Exact bolometric planar breakout solutions}
\author{Nir Sapir\altaffilmark{1}, Boaz Katz\altaffilmark{2,3}, and Eli Waxman\altaffilmark{1}}

\altaffiltext{1}{Dept. of Particle Phys. \& Astrophys., Weizmann Institute of Science, Rehovot 76100, Israel}
\altaffiltext{2}{Inst. for Advanced Study, Princeton, NJ 08540, USA}
\altaffiltext{3}{John Bahcall Fellow, Einstein Fellow}

\begin{abstract}
The problem of a non-steady planar radiation mediated shock (RMS) breaking out from a surface with a power-law density profile, $\rho\propto x^n$, is numerically solved in the approximation of diffusion with constant opacity. For an appropriate choice of time, length and energy scales, determined by the breakout opacity, velocity and density, the solution is universal, i.e. depends only on the density power law index $n$. The resulting luminosity depends weakly on the value of $n$. An approximate analytic solution, based on the self-similar hydrodynamic solutions \citep{Sakurai60} and on the steady RMS solutions \citep[e.g.][]{Weaver76}, is constructed and shown to agree with the numerical solutions as long as the shock is far from the surface, $\tau\gg c/\vt_{\rm sh}$. Approximate analytic expressions, calibrated based on the exact solutions, are provided, that describe the escaping luminosity as a function of time. These results can be used to calculate the bolometric properties of the bursts of radiation produced during supernova (SN) shock breakouts. For completeness, we also use the exact breakout solutions to provide an analytic approximation for the maximum surface temperature for fast ($\vt_{\rm sh}\gtrsim0.1$) non-thermal breakouts, and show that it is few times smaller than inferred based on steady-state RMS solutions.
\end{abstract}

\section{Introduction}
\label{sec:intro}

During a SN explosion, a strong shock wave propagates through and ejects the stellar envelope. This shock moves with high velocities, $\bt_{\rm sh}\equiv\vt_{\rm sh}/c\gtrsim 0.01$, through plasma with densities $\rho\sim 10^{-10}$ - $10^{-6}\gr \cm^{-3}$. Under such conditions the energy density is dominated by radiation, and the opacity is dominated by electron scattering. The kinetic energy is converted to thermal energy through Compton scattering of free electrons. The transition layer of such a Radiation Mediated Shock (RMS) has a finite width, with optical depth of $\delta \tau\sim c/\vt_{\rm sh}$ \citep{Weaver76}.

As long as the shock is deep within the envelope, where the density does not vary significantly over the shock width, its propagation is well described by the (ideal fluid) hydrodynamics equations, considering the shock as a discontinuity. For envelope density that varies as a power-law of the distance from the stellar edge, as is the case for a wide range of progenitors, the hydrodynamic profiles are given by the self similar solutions of \citet{Sakurai60}. Once the shock reaches a distance from the surface which is comparable to the width of the front, i.e. an optical depth $\tau$ from the surface for which $\tau\sim c/\vt_{\rm sh}$, the hydrodynamic solutions are no longer valid. At this point, the radiation in the shock transition region diffuses out and a burst of radiation is emitted \citep[][]{Colgate74b,Falk78,Klein78,Ensman92,Matzner99,Blinnikov00,Katz09,Piro10,Nakar10}. The light curve and spectrum of the escaping radiation depends on the structure of the shock transition.

The structure of steady state RMSs propagating through a homogeneous medium with constant velocity was solved both in the non-relativistic \citep{Weaver76} and in the highly relativistic \citep{BKW10} regimes. An approximate analytic description of the shock structure was derived \citep{Katz09} for fast shocks, $\bt_{\rm sh}\gtrsim0.1$, where the radiation is in Compton equilibrium and far from thermal equilibrium \citep{Weaver76}. These solutions do not provide, however, an accurate description of the shock near breakout, since at breakout the density varies on a scale comparable to the shock width and the shock begins to lose energy to the outflow of radiation.

The emitted burst was previously studied either by numerically calculating the early emission from specific progenitors \citep[including][]{Ensman92,Blinnikov00,Utrobin07,Tominaga09,Tolstov10,Dessart10}, or by providing analytic order of magnitude estimates for the breakout properties based on the steady state solutions \citep[e.g.][]{Matzner99,Katz09,Nakar10}. The bolometric properties of a RMS breaking out from an envelope with an exponential density profile was solved by \citet{Lasher79} (We note that the reported results were restricted to early times, in which only $~0.5$ of the energy is emitted, see \sref{sec:Luminosity}). The accurate time dependent spectrum of shock breakout from a general progenitor has not yet been calculated.

For non-relativistic breakouts, the transport of radiation is well described by diffusion. As argued below, in \S~\ref{sec:Results}, the diffusion approximation holds at least for $\vt_{\rm sh}/c<0.3$. Moreover, for such velocities the maximum electron temperature reached is $\ll m_ec^2$ (see fig.~\ref{fig:beta_vs_Tmax_rho}), which implies that the scattering cross section is frequency independent. Thus, the hydrodynamic and radiation energy density profiles may be obtained by solving the hydrodynamics equations coupled to a constant opacity diffusion equation for the radiation energy density \citep{Lasher79}. In this paper, the time dependent hydrodynamic and radiation flux profiles before, during and following breakout are derived by solving these equations (neglecting plasma pressure).

The exact formulation of the problem is given in \sref{sec:Formulation of the problem}, the numerical solutions are described in \sref{sec:Numerical Integration}, and the results for the hydrodynamic profiles and emitted flux and energy (per unit area) are presented in \sref{sec:Results}. These results are used in the second paper of this series \citep{Katz11} to find the hydrodynamic and bolometric properties of SN shock breakouts taking into account the finite radius of the progenitor and the lowest order relativistic corrections. In \S~\ref{sec:Temperature} we explain how the exact solutions obtained here may be used to derive the spectral properties of the breakout. The temperature profiles and the spectral properties of the breakout are described in a third paper in this series \citep{Sapir11b}. For completeness, we give in \S~\ref{sec:Temperature} the maximum surface temperature obtained for various breakout velocities and densities. Our main conclusions are summarized in \sref{sec:Discussion}.

\section{Formulation of the problem}
\label{sec:Formulation of the problem}

Consider the problem of a planar RMS propagating in the positive $x$ direction through a plasma occupying the $x<0$ region. The initial, $t=-\infty$, density of the plasma decreases towards the initial position of the surface, $x=0$, as
\begin{equation}\label{eq:nDefinition}
\rho|_{t=-\infty}\propto \left\{\begin{array}{cc}
\abs{x}^n&x<0\\
0&x>0
\end{array}\right.\propto \tau^{n/(n+1)}.
\end{equation}
$\tau(x,t)$ is the optical depth of the matter lying ahead of $x$,
\begin{equation}
\tau=\int_{x}^{\infty}\kappa\rho dx=\kappa\int_{x}^{\infty}\rho dx,
\end{equation}
where $\kappa$ is the constant opacity.

The analysis is preformed under the following assumptions/approximations:
\begin{itemize}
\item The velocities are non relativistic, $\bt=\vt/c\ll 1$;
\item The internal energy (and pressure) of the matter are neglected;
\item Photon transport is described by diffusion with a constant opacity $\kappa$.
\end{itemize}

Asymptotically, at large optical depth from the surface, the hydrodynamic profiles are described by the self-similar solutions of \citet{Sakurai60}. The shock velocity follows a power-law dependence on the optical depth $\vt_{\rm sh}\propto\tau^{-\lm(n,\gamma)/(n+1)}$, with $\lm$ depending on the density profile index $n$ and on the value of the adiabatic index $\gamma$. In the case of radiation domination, $\gamma=4/3$ and $\lm(n)\approx0.19n$ \citep{Grasberg81}. In the limit $n\rightarrow \infty$ the problem corresponds to a strong shock in an exponential density profile \citep{Hayes68}.

The shock has a finite width $\de\tau\sim \bt_{\rm sh}^{-1}$ \citep[e.g.][]{Weaver76} where $\bt_{\rm sh}=\vt_{\rm sh}/c$. At an optical depth of $\tau\sim\bt_{\rm sh}^{-1}$, energy diffuses through the surface and the shock dissolves. Note, that at any finite optical depth, the hydrodynamic profile is "smoothed", i.e. the shock transition layer is not a discontinuity, due to the diffusion of energy, and the "shock velocity" is not well defined.

\subsection{Hydrodynamic and energy diffusion equations}
The following Lagrangian coordinates are used to describe the flow. For each mass element, $m=-\tau/\kappa$ is minus the mass per unit area enclosed between the element and the surface while $x,\vt$ and $e$ are the position, velocity and energy density of the element respectively.  In this problem, all mass elements start with negative positions and throughout the evolution move outward with positive velocities.

Under the above assumptions, the equations for $x,\vt$ and $e$ as a function of $m$ and the time $t$ are
\begin{align}
&\pr_t x=\vt, \label{eq:VelocityDef}\\
&\pr_{t}\vt=-\pr_mp,\label{eq:MomentumCons}\\
&\pr_t(e/\rho)=-\pr_mj-p\pr_m\vt,\label{eq:EnergyCons}\\
&j=-\frac{c}{3\kappa}\pr_me,  \label{eq:CurrDiff}\\
&e=3p \label{eq:EOS}
\end{align}
where $\rho=\pr_xm$ is the density, $j$ is the energy flux (energy current density), $p$ is the radiation pressure and $\kappa$ is the constant opacity.

The initial density profile and the asymptotic shock velocity at large optical depth can be parameterized by
\begin{equation}\label{eq:InitialProfileRho}
\rho_{\rm in}=\rho_0 (\bt_0\tau)^{n/(n+1)}
\end{equation}
and
\begin{equation}\label{eq:InitialProfileV}
\bt_{\rm sh}\xra[\tau\ra\infty]{}\bt_0 (\bt_0\tau)^{-\lm/(n+1)}
\end{equation}
respectively. Note that the equations are normalized so that the density and shock velocity at $\tau=\bt_0^{-1}$ would equal $\rho_0$ and $\beta_0$ respectively if the hydrodynamic solution \citep{Sakurai60} was applicable to this point. That is, breakout occurs at $\tau\sim\bt_0^{-1}$ with $\bt_{\rm sh}\sim\bt_0$.

Radiation is assumed to freely escape from the surface. Given that the smallest optical depth over which changes can occur is much larger than unity, $\tau\sim\bt_0^{-1}\gg1$, the energy density $e$ is set to zero at the surface,
\begin{equation}\label{eq:SurfaceBcE}
e(\bt_0\tau=0)=0.
\end{equation}
The velocity at the surface is determined by equations \eqref{eq:VelocityDef}-\eqref{eq:EOS} as follows. Combining equations \eqref{eq:MomentumCons},\eqref{eq:CurrDiff} and \eqref{eq:EOS} the following general relation is obtained \citep{Lasher79}\footnote{Note that Eq. \eqref{eq:AccToCurr1} is a direct consequence of the assumption that the plasma's pressure is negligible. This equation is equivalent to stating that in scattering events of forward/backward symmetry, the momentum of the radiation is completely transferred to the medium. This relation holds independent of the angular distribution of the radiation \citep{Lasher79}.}
\begin{equation}\label{eq:AccToCurr1}
\pr_t\vt=\frac{\kappa}{c}j.
\end{equation}
Since the energy flux is continuous across the surface, the velocity of the surface is related to the flux by \citep{Lasher79}
\begin{equation}\label{eq:SurfaceBcv}
\pr_t\vt(\bt_0\tau=0)=\frac{\kappa}{c}j(\bt_0\tau=0).
\end{equation}

The dimensionful parameters $\kappa,\rho_0$ and  $\vt_0=\bt_0c$ along with the dimensionless parameter $n$ completely define the problem through the initial density profile \eqref{eq:InitialProfileRho}, the asymptotic velocity at early times \eqref{eq:InitialProfileV} and the boundary conditions \eqref{eq:SurfaceBcE} and \eqref{eq:SurfaceBcv}, along with the equations of motion \eqref{eq:VelocityDef}-\eqref{eq:EOS}.

\subsection{Dimensionless equations}
By introducing dimensionless variables, $\tilde A=A/A_0$ with
\begin{equation}\label{eq:A_0}
x_0=\frac{1}{\kappa\rho_0\bt_0},\quad  m_0=\frac{1}{\kappa\bt_0},\quad t_0=\frac{x_0}{\vt_0},\quad p_0=\rho_0\vt_0^2,
\end{equation}
the following dimensionless equations are obtained:
\begin{align}\label{eq:AConsParDimensionless}
\tilde \vt=&\pr_ {\tilde t} \tilde x,\cr
\pr_{\tilde t} \tilde \vt=&-\pr_ {\tilde m} \tilde p,\cr
\pr_{\tilde t}(3\tilde p/ \tilde \rho)=&\pr^2_{\tilde m} \tilde p- \tilde p\pr_{\tilde m} \tilde \vt,
\end{align}
with an initial density profile
\begin{equation}\label{eq:BCInDimensionless}
\tilde \rho_{\rm in}=\abs{\tilde m}^{n/(n+1)},
\end{equation}
asymptotic shock velocity
\begin{equation}\label{eq:AssymptoticVDimensionless}
\tilde \vt_{\rm sh} \xra[\tilde m\ra\infty]{}\abs{\tilde m}^{-\lm/(n+1)},
\end{equation}
and surface boundary conditions,
\begin{equation}\label{eq:BCScDimensionless}
\tilde p(0)=0,~~~~~\pr_{\tilde t}\tilde\vt(0)=-\pr_{\tilde m}\tilde p|_{\tilde m=0}.
\end{equation}
The system of equations \eqref{eq:AConsParDimensionless}-\eqref{eq:BCScDimensionless} has one free parameter, namely the value of the initial density power-law index $n$. Therefore, for a given value of $n$, the time dependent profiles of the dimensionless quantities $\tilde\vt,\tilde p,\tilde x,\tilde \rho$ are universal (independent of $\vt_0$, $\rho_0$ and $\kappa$).  This universality is not surprising.  For a given value of $n$, the (dimensionful) problem is defined by the values of three dimensionful parameters, namely $\vt_0$, $\rho_0$ and $\kappa$ as well as the speed of light $c$. The only dimensionless parameter (aside from $n$) that can be obtained is $\vt/c$, which is taken to zero in the non relativistic limit.

\section{Numerical Integration}\label{sec:Numerical Integration}
Equations \eqref{eq:AConsParDimensionless} were numerically solved using an implicit scheme in time, a grid with uniform spacing in mass, and using constant time steps.
Any boundary condition on the inner boundary, combined with an initial power law density profile would result in an accelerating shock which would approach the Sakurai solution, \eqref{eq:AssymptoticVDimensionless}, far from the boundary,  as long as the box is large enough. In order to achieve fast convergence with box size,  the initial distributions of positions, velocities and pressure were chosen to be an approximation of the solution at an early stage when the shock was close to the inner boundary. The precise profiles were described by an ansatz combining the steady state structure of a radiation mediated shock with the time dependent Sakurai self-similar solution (see \sref{sec:SakuraiWeaverAnsatz}). The velocity and energy flux of the inner boundary (as a function of time) were set to be equal to the values given by the Sakurai solution. The resulting profiles converged quickly with box size as shown below.

As a consistency check, a set of different boundary conditions was used in an additional calculation. In this case, the inner boundary moves with a constant velocity and energy is deposited at the innermost cell at the initial time. The resulting profiles were found to converge to the same profiles that were obtained with the Sakurai-Weaver ansatz, albeit much slower.

The hydrodynamic profiles shown below were obtained with a 6000 cells box of width $d\tilde m= 0.08$ and time steps of $d\tilde t\approx 5\times 10^{-5}$. Initially, the shock was positioned at a (mass) distance of $\de \tilde m\approx 10$ from the inner boundary. The emitted energy flux and fluence results were obtained with a 1200 cells box of width $d\tilde m= 0.025$ and time steps of $d\tilde t\approx 1\times 10^{-4}$. Table \ref{tab:convergenceTable} presents the convergence with increasing resolution and box size of parameters describing the characteristics of the flux and fluence (see \ref{sec:Luminosity} for definitions). As indicated by the table, numerical results presented for the values of these parameters are converged to an accuracy of better then $1\%$. The table presents convergence with respect to box size and resolution. The convergence with respect to changes in time step size is better than $10^{-5}$ and not presented.

\begin{table} \begin{tabular}{|c|cc|cc|}
\hline
& $\Delta \mathcal{L}_{\rm peak}/\mathcal{L}_{\rm peak}$ \footnotemark[1]& $\Delta t_{\rm peak}/t_{\rm peak}$ \footnotemark[1] & $\Delta \mathcal{E}_{\infty}/\mathcal{E}_{\infty}$\footnotemark[1] & $\Delta a_t/a_t$\footnotemark[1]\tabularnewline
\hline
box size /2\footnotemark[2] & $1.1\times10^{-2}$ & $1.2\times 10^{-2}$ & $4.8\times 10^{-3}$ & $2.3\times 10^{-3}$ \\
box size $\times$ 2\footnotemark[2] & $1\times 10^{-3}$ & $5\times 10^{-3}$ & $4\times 10^{-4}$ & $1.2\times 10^{-2}$ \\
\hline
resolution /2\footnotemark[2] & $1.4\times 10^{-2}$ &
$2.8\times 10^{-2}$ & $2.3\times 10^{-4}$ & $8.6\times 10^{-4}$ \\
 resolution $\times 2$\footnotemark[2]& $3\times 10^{-3}$ & $1.3\times 10^{-2}$ & $7\times 10^{-5}$ & $5.5\times 10^{-4}$ \\
\hline
\end{tabular}
\footnotetext[1]{See table \ref{tab:luminosityTable} for definition and resulting value}
\footnotetext[2]{With respect to the nominal calculation described in the text}
\caption{Convergence of the numerical solution\label{tab:convergenceTable}}
\end{table}

\section{Results}\label{sec:Results}
The resulting profiles of the density, velocity and pressure as a function of normalized mass at different times are given in \sref{sec:Profiles} for $n=3$ . The shock velocity at different times is calculated using different prescriptions, and compared to the hydrodynamic self-similar velocity, in \sref{sec:ShockVelocity}. The resulting emission of radiation from the surface and the asymptotic distribution of velocities is presented in \sref{sec:Luminosity}. $t/t_0=0$ is chosen as the time when the Sakurai shock reaches the surface, according to the integration of \eqref{eq:InitialProfileV}.

\subsection{Density, velocity and pressure profiles}\label{sec:Profiles}
\begin{figure}[h]
\epsscale{1} \plotone{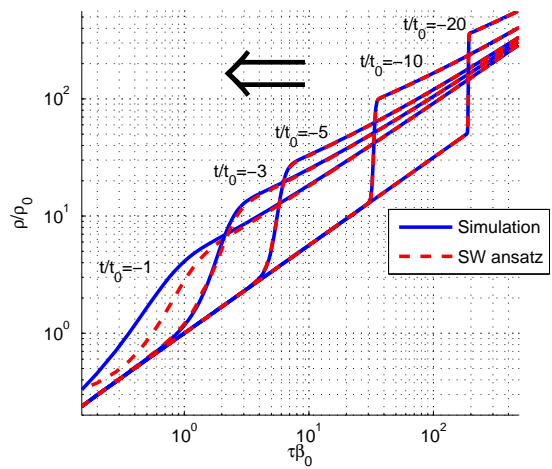}
\caption{The normalized density profiles as a function of the normalized mass (optical depth) at different times prior to breakout (blue solid lines). Also plotted are the density profiles of the \SWA\ at the appropriate times (red dashed lines, see \sref{sec:SakuraiWeaverAnsatz}). The arrow indicated the shock propagation direction. $\beta_0c$ is the shock velocity "at breakout", i.e. the velocity for which $\tau=1/\beta$ in the Sakurai solution, and $t_0$ is defined in eq.~\ref{eq:A_0}. The peak emitted flux is obtained at $t_{\rm peak}/t_0=-1.25$.\label{fig:rho_vs_m}}
\end{figure}

\begin{figure}[h]
\epsscale{1} \plotone{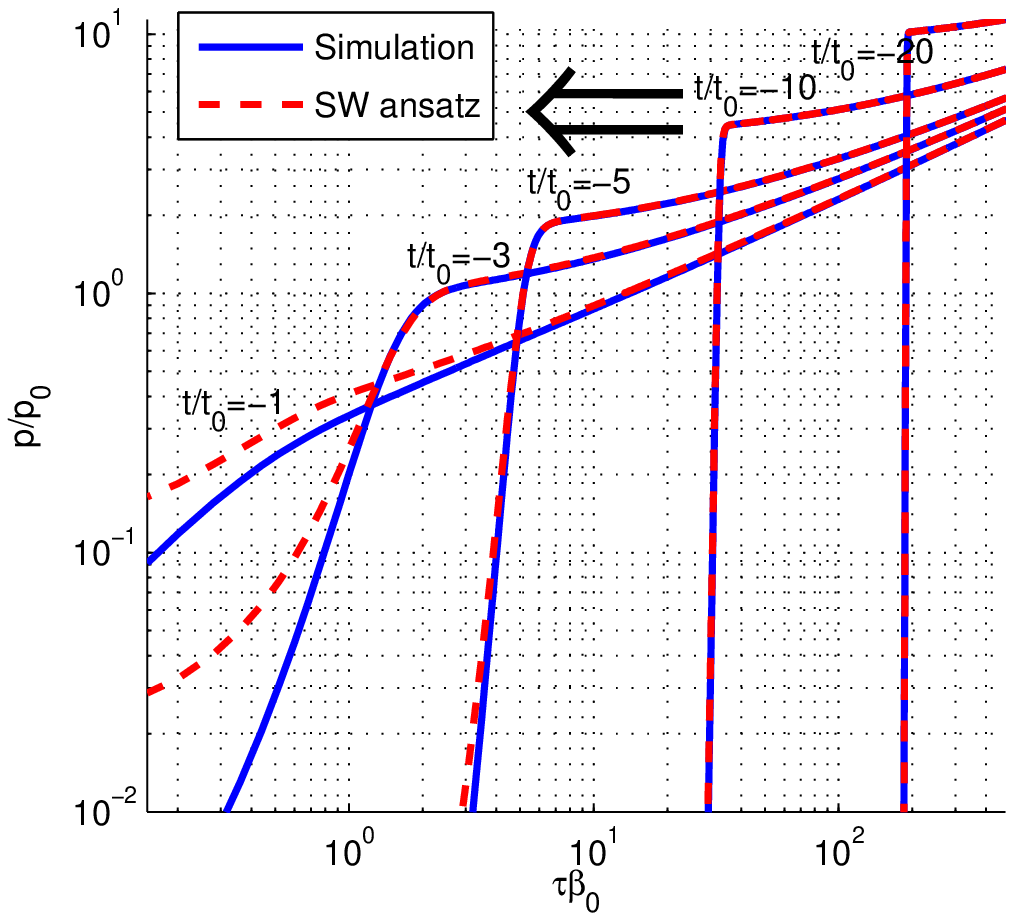}
\caption{The normalized pressure profiles as a function of the normalized mass at different times prior to breakout. Line styles/colors are as in figure \ref{fig:rho_vs_m}.\label{fig:p_vs_m}}
\end{figure}

\begin{figure}[h]
\epsscale{1} \plotone{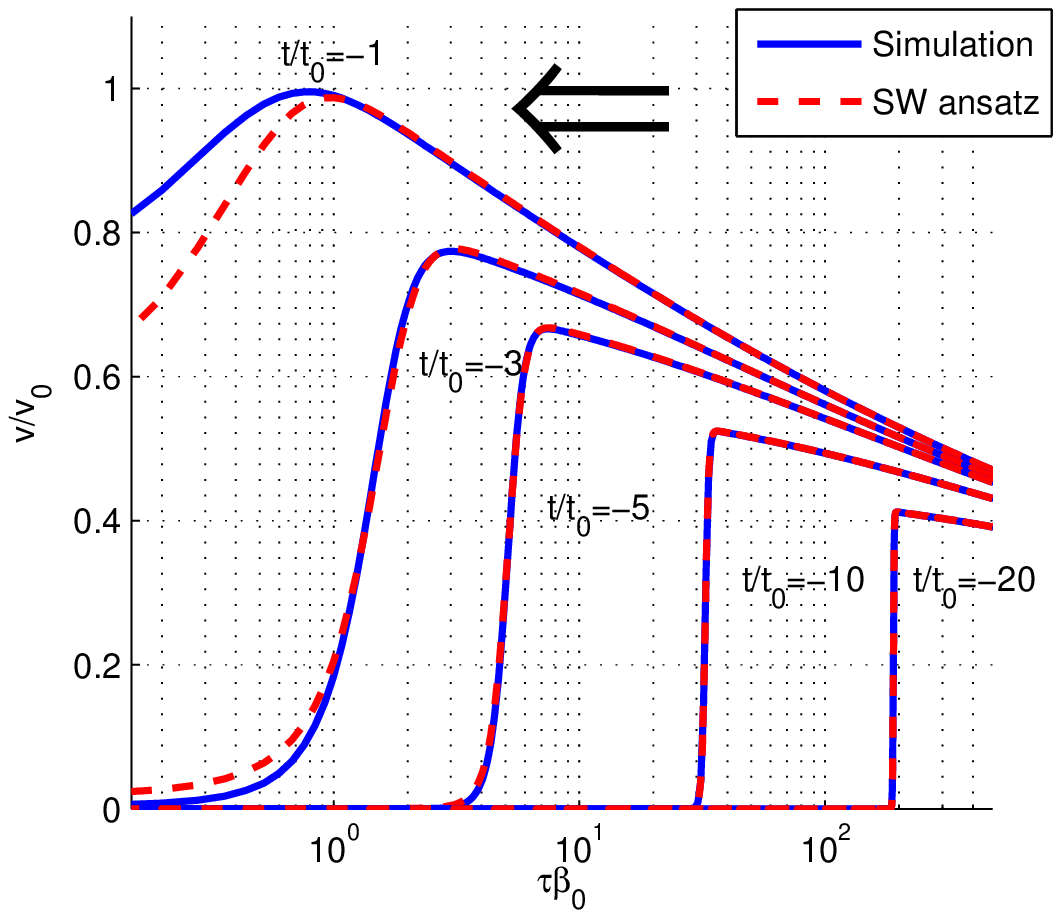}
\caption{The normalized velocity profiles as a function of the normalized mass at different times prior to breakout. Line styles/colors are as in figure \ref{fig:rho_vs_m}.\label{fig:v_vs_m}}
\end{figure}

The hydrodynamic profiles of the density, pressure and velocity of the gas as the shock progresses to the surface are shown in figures \ref{fig:rho_vs_m},  \ref{fig:p_vs_m} and \ref{fig:v_vs_m} respectively. The results of the simulation are compared to a Sakurai-Weaver ansatz that combines the Sakurai accelerating shock with the Weaver steady state RMS, without any fitted parameters (see Appendix). As can be seen, the ansatz agrees with the results to a high accuracy up to the time when the shock reaches an optical depth of $\tau\sim \bt_0^{-1}$.

\begin{figure}[h]
\epsscale{1} \plotone{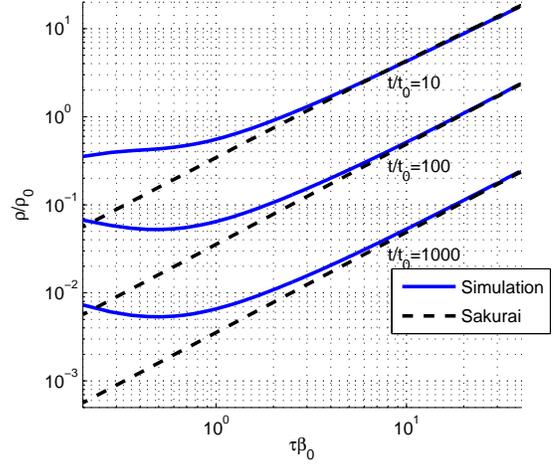}
\caption{The normalized density profiles as a function of the normalized mass at different times following breakout (blue solid lines). Also plotted are the density profiles of the self-similar hydrodynamic Sakurai solution at the appropriate times (black dashed lines).
\label{fig:rho_vs_m_late}}
\end{figure}

\begin{figure}[h]
\epsscale{1} \plotone{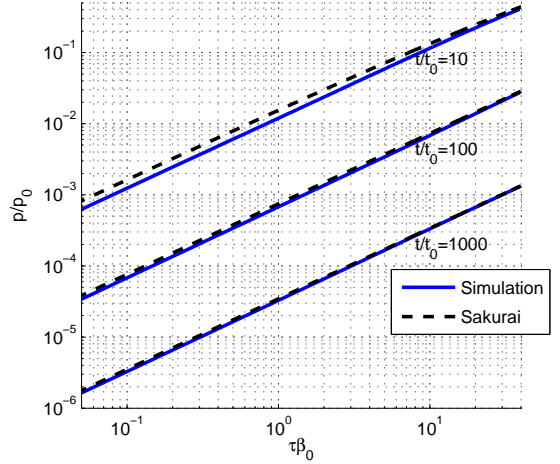}
\caption{Normalized pressure as a function of the normalized mass at different times following breakout. Line styles/colors are as in figure
\ref{fig:rho_vs_m_late}
\label{fig:p_vs_m_late}}
\end{figure}

\begin{figure}[h]
\epsscale{1} \plotone{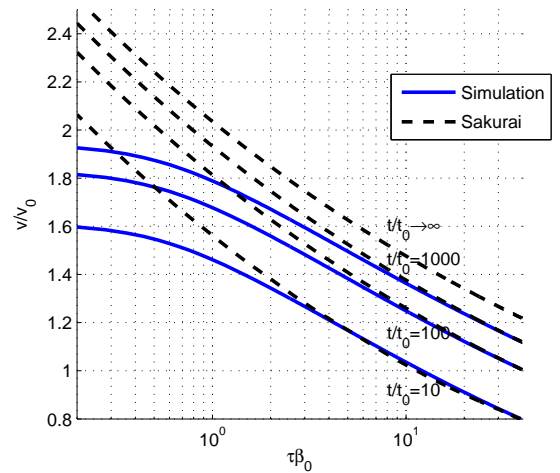}
\caption{Normalized  velocity as a function of the normalized mass at different times following breakout. Line styles/colors are as in figure
\ref{fig:rho_vs_m_late}
\label{fig:v_vs_m_late}}
\end{figure}

The profiles at different times after breakout are plotted in figures \ref{fig:rho_vs_m_late}-\ref{fig:v_vs_m_late}. The results of the simulation are compared to the Sakurai solution of the pure hydrodynamic equations. Note that the \SWA\ does not apply at these times, which are greater than the breakout time, since a shock no longer exists. As can be seen in the figures, diffusion affects the profiles only close to the boundary  $\tau\lesssim\beta_0^{-1}$. This is expected since the diffusion depth has a constant optical depth  $\tau_{\rm diff}\sim\beta_0^{-1}$ throughout the expansion \citep[e.g.][]{Matzner99, Piro10, Nakar10}.   The difference between the Sakurai solution and the numerical solution at the point $\tau=\beta_0^{-1}$ is $\sim10\%$.  Note that in the case $n=0$, the optical depth of the diffusion depth grows with time and the deviations from the Sakurai solution are observed on this growing scale.

The results of the velocity profiles also show for which velocities the diffusion approximation is appropriate in describing the radiation transport. The diffusion approximation is valid as long as the velocity does not change significantly over a photon mean free path, and this requirement can be expressed as $\pr_\tau\bt\ll 1$. During propagation, the velocity derivative increases as the shock approaches the surface and the maximal value is $\pr_\tau\bt\approx1.0\bt_0^2$. Therefore, we conclude that the diffusion approximation is likely valid for $\bt_0\lesssim0.3$.

\subsection{Shock Velocity}\label{sec:ShockVelocity}
Strictly speaking, the position and the velocity of the shock are not well defined, since the diffusion of radiation smooths the hydrodynamic shock discontinuity. Here, different prescriptions for defining the shock velocity given the hydrodynamic profiles are used for comparing the results to the pure hydrodynamic Sakurai solution (which has a well defined shock position and velocity).
The effective shock position $\msheff(t)$ is defined as the point where the density increased by some chosen factor $1<f<7$ compared to the original density,
\begin{equation}
\rho(\msheff,t)=f\rho_{\rm in}(\msheff).
\end{equation}
In this section, $f$ is chosen to be $f=\sqrt{7}$. The effective shock velocity is obtained by the following two prescriptions which reduce to the correct shock velocity in the steady state case.
\begin{enumerate}
\item The effective shock velocity is set to be the time derivative of the effective shock position,
\begin{equation}\label{eq:vsheffa}
\vsheff^a(t)\equiv \frac{d}{dt}x(\msheff,t).
 \end{equation}
\item Based on the steady state shock relation in the upstream frame, $p=\rho_{\rm up}\vt\vt_{\rm sh}$ (which holds for a steady state shock moving with velocity $\vt_{\rm sh}$ through a homogeneous medium with $\rho_{\rm up}$), the effective shock velocity is set to be
\begin{equation}\label{eq:vsheffb}
\vsheff^b(t)\equiv \frac{p(\msheff,t)}{\rho_{\rm in}(\msheff)\vt(\msheff,t)}.
\end{equation}
\end{enumerate}

\begin{figure}[h]
\epsscale{1} \plotone{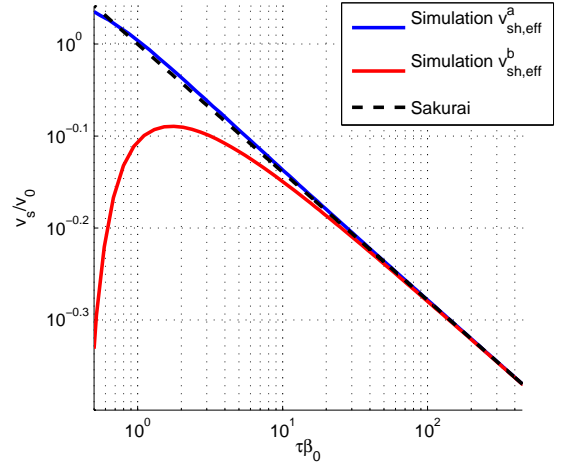}
\caption{The normalized effective shock velocities as a function of the effective position of the shock (in terms of normalized mass). The blue and red solid lines are the effective shock velocities estimated from equations \eqref{eq:vsheffa} and \eqref{eq:vsheffb} respectively and the black dashed line is the Sakurai solution. The separation of the red and blue curves is an indication of the deviation from the (ideal fluid) hydrodynamic evolution. \label{fig:betas_vs_m}}
\end{figure}

The numerical shock velocity is compared in figure \ref{fig:betas_vs_m} to that obtained in the self-similar (ideal fluid) hydrodynamic solution. As can be seen in the figure, when the shock is far from the surface, the two definitions of the effective shock velocity agree with each other and with the Sakurai solution.
The two definitions lead to different estimates of shock velocity close to the surface. This is an indication of the deviation from the (ideal fluid) hydrodynamic evolution, which is expected to occur when the shock width is comparable to the distance from the surface.

\subsection{Emission of radiation from the surface}\label{sec:Luminosity}
\begin{figure}[h!]
\epsscale{1} \plotone{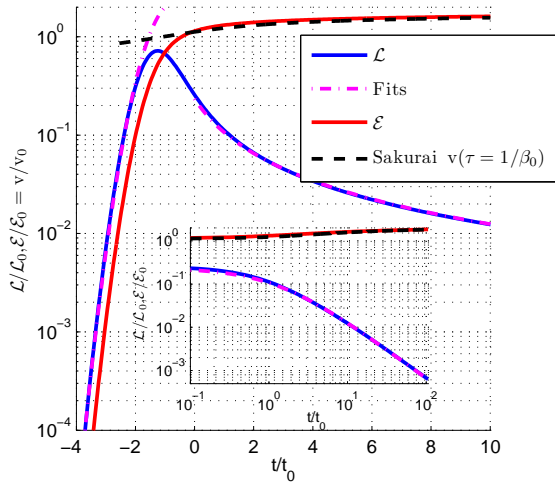}
\caption{Normalized (see eq.~\ref{eq:LE_norm}) emitted energy flux (solid blue line) and energy per unit area (solid red line) as a function of time since (expected) breakout. These are equal to the normalized acceleration and velocity of the outermost mass element (see discussion in \sref{sec:Luminosity}).  Also plotted are the fits to the emitted flux given by Eq. \eqref{eq:ELApprox} for $t/t_0>0$ and by Eq. \eqref{eq:earlyL} for $t/t_0<-1$ (magenta dashed-dotted lines) and the velocity of the $\tau\bt_0=1$ mass element in the Sakurai solution (dashed black line). The inset shows the same quantities on a larger time scale.\label{fig:L_vs_t}}
\end{figure}

We next consider the energy flux (i.e. luminosity per unit area) emitted from the surface, $\mathcal{L}(t)=j(\bt_0\tau=0,t)$, and the amount of energy per unit area emitted up to a given time $t$
\begin{equation}
\mathcal{E}(t)=\int_{-\infty}^{t}\mathcal{L}(t')dt',
\end{equation}
during and following the shock breakout. The emitted flux and energy, normalized to
\begin{equation}\label{eq:LE_norm}
    \mathcal{L}_0=\rho_0\vt_0^3,\quad \mathcal{E}_0=\kappa^{-1}\beta_0c^{2},
\end{equation}
are shown in figure \ref{fig:L_vs_t}. As can be seen in the figure, the emitted flux rises sharply and later decreases, asymptotically approaching a power law $\mathcal{L}(t)\propto t^{-4/3}$. The amplitude and position of the peak and the full width at half maximum are given in table \ref{tab:luminosityTable}.

The late time power law decline, $\mathcal{L}(t)\propto t^{-4/3}$, is expected for any density profile index $n>0$  \citep{Matzner99,Piro10,Nakar10} . To see this note that the diffusion optical depth is constant in time and the luminosity is proportional to the pressure at $\tau\sim\bt^{-1}$ which roughly decreases adiabatically $p\propto \rho^{4/3}$ with the density  decreasing linearly with time. Note that in the case $n=0$ the diffusion optical depth grows with time. Taking this into account, the asymptotic emitted flux follows $\mathcal{L}(t)\propto t^{-9/8}$ for this case.
The dimensionless emitted energy and energy flux can be approximated at late time ($t>t_{\rm peak}+a_t t_0$) by
\begin{align}\label{eq:ELApprox}
&\mathcal{E}(t)=\mathcal{E}_{\infty}\left[1-\left(\frac{t-t_{\rm peak}}{a_t t_0}\right)^{-1/3}\right]\cdot(t>t_{\rm peak}+a_tt_0),\cr
&\mathcal{L}(t)=\frac{\mathcal{E}_{\infty}}{3a_t t_0}\left(\frac{t-t_{\rm peak}}{a_t t_0}\right)^{-4/3}\cdot(t>t_{\rm peak}+a_tt_0),\cr
\end{align}
where $t_{\rm peak}$ is the time at which $\mathcal{L}(t)$ reaches its maximum, $a_t$ is a free parameter, and $(x>y)$ is a function that is unity for $x>y$ and zero otherwise.  The fitted values of $a_t$ and $\mathcal{E}_{\infty}$ for $n=3/2,3$ are given in table \ref{tab:luminosityTable}. The corresponding power-law fit for the case $n=3$ is shown in figure \ref{fig:L_vs_t}.

\begin{table*}
\begin{tabular}{|c|ccc|cc|ccc|}
\hline
 &\multicolumn{3}{|c|}{Peak characteristics}&\multicolumn{2}{|c|}{Eq. \eqref{eq:ELApprox} Fit}&\multicolumn{3}{|c|}{Eq. \eqref{eq:earlyL} Fit}\\
\hline
$n\footnotemark[1]$ & $\mathcal{L}_{\rm peak}\footnotemark[2]/\mathcal{L}_0$ & $t_{\rm peak}\footnotemark[3]/t_0$  & $\Delta t_{FWHM}/t_0\footnotemark[4]$ & $\mathcal{E}_{\infty}\footnotemark[5]/\mathcal{E}_0=\vt_{\infty}/\vt_0$& $a_t$\footnotemark[5]&$\mathcal{L}_i/\mathcal{L}_0$ \footnotemark[6] & $a_i$ \footnotemark[6] & $b_i$ \footnotemark[6]\tabularnewline
\hline
3  & 0.72 & -1.25 & 1.61 & 2.03 & 0.1 & 1.25 & 1.11 & 1.57 \tabularnewline
3/2 & 0.77 &  -0.78 & 1.49 & 2.13 & 0.129 & 0.45 & 1.08 & 1.14 \tabularnewline
\hline
\end{tabular}
\footnotetext[1]{Density power law index, Eq. \eqref{eq:nDefinition}}
\footnotetext[2]{Peak Luminosity}
\footnotetext[3]{Time of peak luminosity, measured from time of expected breakout}
\footnotetext[4]{Luminosity full width half maximum}
\footnotetext[5]{See Eq. \eqref{eq:ELApprox}}
\footnotetext[6]{See Eq. \eqref{eq:earlyL}}
\caption{Light curve characteristic values\label{tab:luminosityTable}}
\end{table*}

The early stages of the rise in emitted energy flux are due to photons diffusing ahead of the shock while it is still far from the surface. Since the distribution of these photons drops like $\propto e^{- x^2}$ away from the shock, the rise in the emitted energy flux can be approximately described by
\begin{equation}\label{eq:earlyL}
\mathcal{L}(t)=\mathcal{L}_{\rm i}e^{-a_{\rm i}(t/t_0)^2-b_{\rm i}(t/t0)}.
\end{equation}
The fitted values of $\mathcal{L}_{\rm i},a_{\rm i}$ and $b_{\rm i}$ are given in table \ref{tab:luminosityTable} and the corresponding fit for the case $n=3$ is shown in figure \ref{fig:L_vs_t}.

Using equation \eqref{eq:SurfaceBcv}, the acceleration and velocity of the surface (outermost mass element) are related to the emitted energy flux and fluence by
\begin{equation}
\pr_t\vt(\bt_0\tau=0,t)=\frac{\kappa}{c} \mathcal{L}(t)
\end{equation}
and
\begin{equation}\label{eq:Velocity_Energy}
\vt(\bt_0\tau=0,t)=\frac{\kappa}{c} \mathcal{E}(t)
\end{equation}
respectively.
Thus the plotted normalized emitted flux and fluence are, respectively, also the normalized acceleration and velocity of the surface
\begin{align}
&\pr_t\vt(\bt_0\tau=0,t)t_0/\vt_0=\mathcal{L}(t)/\mathcal{L}_0,\cr
&\vt(\bt_0\tau=0,t)/\vt_0=\mathcal{E}(t)/\mathcal{E}_0,
\end{align}
and the surface velocity can be approximated by
\begin{align}
\vt(\bt_0\tau=0,t)\approx  \vt_{\infty}\left[1-\left(\frac{t-\tref}{a_t t_0}\right)^{-1/3}\right]\cdot(t>\tref)
\end{align}
where $\vt_{\infty}/\vt_0=\mathcal{E}_{\infty}/\mathcal{E}_0$ is given in table \ref{tab:luminosityTable}.

For comparison, the velocity of the mass element at $\tau=\beta_0^{-1}$ in the pure hydrodynamic Sakurai solution is also plotted in figure \ref{fig:L_vs_t} (dashed dotted lines, Note that the velocity of the surface in the Sakurai solution is $0$ for $t<0$ and $\infty$ at $t>0$).  As can be seen, the pure hydro solution can be used to approximate the emitted radiation and the surface motion at late times. In particular, the asymptotic velocity achieved by the surface is equal to the maximal velocity achieved in the Sakurai solution (for the mass element at $\tau=\beta_0^{-1}$) to an accuracy of $\sim1\%$. This implies that previous estimates of the velocity of the fastest element \citep{Matzner99} are accurate. As far as we can tell, this high level of accuracy is coincidental.

\section{Temperature}\label{sec:Temperature}
\begin{figure}[h]
\epsscale{1} \plotone{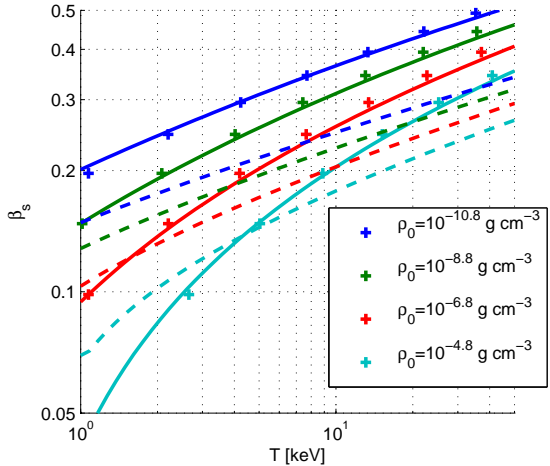}
\caption{The shock velocity $\bt_0$ as a function of maximum surface temperature for different values of breakout density $\rho_0$. Crosses represent the calculated value, solid curves show the curved fit and dashed curves show the results of equation (18) of \citet{Katz09} for steady-state RMS. \label{fig:beta_vs_Tmax_rho}}
\end{figure}

In a following paper \citep{Sapir11b}, the spectral components of the breakout burst are calculated by extending the planar calculation to include a determination of the temperature. This is done by solving a second diffusion equation for the photon number density. Assuming Compton equilibrium, the temperature can be obtained from the pressure and photon number density. In particular, this method allows the first accurate calculation of the spectra for fast shocks $\bt\gtrsim 0.1$ where the radiation is far from equilibrium \citep[][]{Weaver76,Katz09,Nakar10}. This can be compared with the analytical, order of magnitude estimates of peak temperature \citep[][]{Katz09,Nakar10}, and expected spectral features \citep{Nakar10}. For illustration, the relation between peak temperature $T_{\max}$, $\rho_0$ and $\vt_0$ is shown in figure \ref{fig:beta_vs_Tmax_rho} for breakout in an envelope consisting of pure hydrogen or pure helium (the results depend on the atomic number and charge through the combination $Z^2/A$). Also plotted is the prediction of equation (18) in \citet{Katz09} for steady state RMS. The resulting temperature deviates from the value achieved by a steady state shock propagating in a density $\rho_0$ with velocity $\vt_0$, used as an approximation for the break out temperature \cite[][]{Katz09,Nakar10} by factors of a few.

The relation between $\bt_0$, $\rho_0$ and $T_{\max}$ can be fitted by,
\begin{equation}\label{eq:beta_vs_T_rho_relation}
\beta_s=a\log_{10}^2(T_{\rm {keV}})+b\log_{10}(T_{\rm keV})+c\log_{10}(\rho_{-9})+d
\end{equation}
where $T=T_{\rm {keV}}~\rm {keV}$ and $\rho=10^{-9}\rho_{-9}~\gr \cm^{-3}$. For $n=3$, relevant to BSGs and WRs where the velocities can reach $>0.1c$, the best fit parameters are $a=0.03$, $b=0.133$ , $c=-0.0267$ and $d=0.153$. The photon energies at the spectral peak are approximately $h\nu\approx
3T_{\max}$.  The detailed properties of the spectrum of the emitted burst will be studied in \citep[][]{Sapir11b}.

\section{Discussion}\label{sec:Discussion}
The problem of a planar RMS breaking out of a gas with a power-law density profile and with constant opacity was solved in the diffusion approximation. The main results are summarized below.
\begin{itemize}
\item For a given density power law index $n$ the emitted energy flux and the hydrodynamic behavior have universal profiles in space and time, up to scaling of density, velocity and opacity. The scaling of time, distance, mass and energy density are given in Eq.~\eqref{eq:A_0} (in this equation, $\beta_0c$ is the shock velocity "at breakout", i.e. the velocity for which $\tau=1/\beta_{\rm sh}$ in the Sakurai solution, and $\rho_0$ is the initial density at $\tau=1/\beta_0$).
\item The emitted energy flux can be fitted by a super exponential rise, $\mathcal{L}\propto e^{-t^2}$, at early times and by a power law decay, $\mathcal{L}\propto t^{-4/3}$, at late times, see Eqs.~\eqref{eq:earlyL} and \eqref{eq:ELApprox} (for $n=0$ the late time luminosity follows $\mathcal{L}\propto t^{-9/8}$). The values of the parameters that appear in these eqs., along with properties of the emitted flux peak, are provided in table \ref{tab:luminosityTable} (the normalization is given in eq.~\ref{eq:LE_norm}). The emitted energy flux as a function of time for power law indexes in the range $n=1-10$ is fully provided in table \ref{tab:L_of_t} and shown in figure \ref{fig:L_n_vs_t}.
\item The luminosity depends weakly on the decreasing density structure (e.g. on the value of $n$). For power law indexes in the range $n=1-10$ the luminosity changes by less than $25\%$ (see table \ref{tab:L_of_t} and figure \ref{fig:L_n_vs_t}).
\item The acceleration (velocity) of the different mass elements is linearly related to the energy flux (fluence) passing through the mass element by
$\pr_t\vt=\kappa/c j$ \citep[see Eq. \eqref{eq:AccToCurr1}, and][]{Lasher79}. In particular the velocity of the surface, which is also the fastest moving element, is given by the fluence emitted from the surface.
\item At late times, $t\gtrsim 2t_0$, the emitted energy flux and fluence are approximately equal to the acceleration and velocity at $\tau=c/\vt_0$ in the Sakurai solution (see figure \ref{fig:L_vs_t}).
\item The hydrodynamic profiles before breakout $t\lesssim -3t_0$ (see figures \ref{fig:rho_vs_m}, \ref{fig:p_vs_m} and \ref{fig:v_vs_m}) can be accurately described by an ansatz (see \sref {sec:SakuraiWeaverAnsatz}) combining the hydrodynamic power-law solutions of Sakurai and RMS steady state solutions of Weaver.
\item There is no density discontinuity at late stages, suggesting that a collisionless shock does not form in a planar breakout. The formation of the dense shell observed in \citep{Ensman92}, if real, must be related to other physical processes  (e.g. spherical expansion or transport of radiation at $\tau\lesssim1$).
\item We explained in \S~\ref{sec:Temperature} how the exact solutions obtained here may be used to derive the spectral properties of the breakout. The temperature profiles and the spectral properties of the breakout are described in detail in \citep{Sapir11b}. For completeness, we gave in \S~\ref{sec:Temperature} the maximum surface temperature obtained for various breakout velocities and densities, see fig.~\ref{fig:beta_vs_Tmax_rho} and eq.~(\ref{eq:beta_vs_T_rho_relation}). For fast ($\vt_{\rm sh}\gtrsim0.1$) non-thermal breakouts the temperature is few times smaller than inferred based on steady-state RMS solutions.
\end{itemize}

\acknowledgments This research was partially supported by ISF, Minerva and Universities Planning \& Budgeting Committee grants.  B.K is supported by NASA through Einstein Postdoctoral Fellowship awarded by the Chandra X-ray Center, which is operated by the Smithsonian Astrophysical Observatory for NASA under contract NAS8-03060.

\appendix

\section{Profile ansatz combining a time dependent hydrodynamic solution and the steady state RMS solution}\label{sec:SakuraiWeaverAnsatz}
In this section we describe an ansatz that approximates the profile of an RMS propagating through a cold non homogenous medium with initial density profile $\rho_{\rm in}(m)$. We found this ansatz to accurately describe the exact profiles for the case of an initial power law density distribution (see figures \ref{fig:rho_vs_m}-\ref{fig:v_vs_m}).

It is assumed that the solution of the pure hydrodynamic problem of an infinitely thin, strong RMS propagating through the medium is known. Velocities are measured in the laboratory frame, where the unperturbed medium is at rest. The hydrodynamic solution is described by the shock position and velocity as a function of time, $m_{\rm sh}(t)$ and $\vt_{\rm sh}$ respectively, and by the values of the position, pressure and velocity of every mass element $m$, $x_H(m,t), p_H(m,t),\vt_H(m,t)$ respectively. The density is given by $\rho_H(m,t)=\pr_{x_H}m$. Denote the unperturbed density immediately upstream of the shock by $\rho_{\s}(t)$ ($\rho_{\s}=\rho_{\rm in}(m_{\s}(t)$). By assumption, for $m>m_{\rm sh}$ we have $p_H(m,t)=\vt_H(m,t)=0$. The ansatz is given by
\begin{align}\label{eq:Ansatz1}
&x_{\rm anz}(m,t)=x_H+(\kappa\rho_{\s}\bt_{\s})^{-1}\hat x(\hat m),\cr
&\vt_{\rm anz}(m,t)=\left\{ \begin{array}{ll}
\vt_H\cdot \hat \vt(\hat m)&m<m_{\s}\\
\frac{6}{7}\vt_{\s}\hat \vt(\hat m)& m>m_{\s}
\end{array}\right.,\cr
&p_{\rm anz}(m,t)=\left\{ \begin{array}{ll}
p_H\cdot \hat \vt(\hat m)&m<m_{\s}\\
\frac{6}{7}\rho_{\s}\vt_{\s}^2\hat \vt(\hat m)& m>m_{\s}
\end{array}\right.,\cr
\end{align}
where \be\hat m=\kappa\bt_{\s}(m-m_{\s})\ee and $\hat x, \hat \vt$ are given by
\begin{align}\label{eq:Ansatz2}
&\hat x=\frac27\ln\left(\frac{1+e^{3\hat m}}2\right)-\frac67\hat m\cdot(\hat m>0),\cr
&\hat\vt=\inv{1+e^{3\hat m}}.\cr
\end{align}

Eq. \eqref{eq:Ansatz1} is motivated by the fact that the spatial width of the shock is $\delta x\sim (\kappa\rho_{\s}\bt_{\s})^{-1}$ and the mass within the shock region is $\de m\sim (\kappa\bt_{\s})^{-1}$. The expressions for $\hat x$ and $\hat \vt$ where uniquely determined by the following requirements: 1. for a steady state shock propagating through a homogenous medium, this ansatz should exactly reduce to the known analytical solution, 2. at the shock and far from the shock the spatial deformation goes to zero
\begin{align}
\hat x\xra[\de\hat m\ra \pm\infty,0]{}0.
\end{align}

A comparison of this ansatz to the exact numerical results for the case of an initial density distribution $\rho\propto \abs{m}^{3/4}$ ($n=3$) is presented in figures \ref{fig:rho_vs_m}-\ref{fig:v_vs_m}.

\section{Numerical values of $\mathcal{L}(t)$.}\label{sec:Lt}
\begin{figure}[h]
\epsscale{1} \plotone{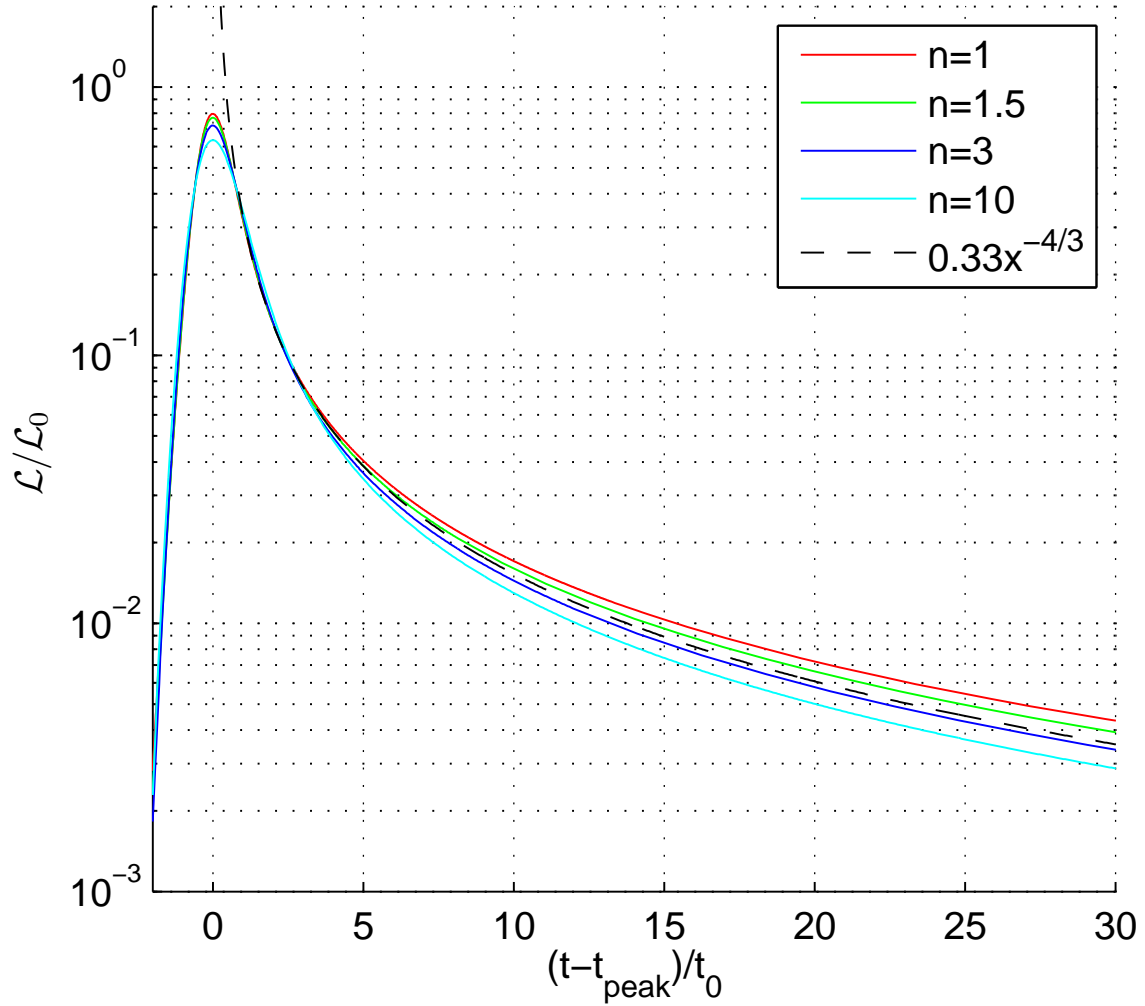}
\caption{Normalized emitted energy flux as a function of the normalized time relative to time of peak emitted energy flux, for different values of the power-law index $n$. Also plotted is the fitting function $\mathcal{L}(t)=0.33\mathcal{L}_0\left[\left(t-t_{\rm peak}\right)/t_0\right]^{-4/3}$ (dashed black line) corresponding to \eqref{eq:ELApprox}.\label{fig:L_n_vs_t}}
\end{figure}

The numerical results for the instantaneous luminosity per unit area, $\mathcal{L}(t)$, resulting from the numerical planar solution,  are given in table \ref{tab:L_of_t} and are plotted in figure \ref{fig:L_n_vs_t} . Time is measured with respect to the time of peak luminosity. The differences in luminosity between the $n=1-10$ cases and the $n=3$ case are less than about $25\%$ at the time interval $-1<(t-t_{\rm peak})/t_0<20$, where about $80\%$ of the energy is emitted.

\begin{table*}[h]
\begin{tabular}{lllll}
$(t-t_{\rm peak})/t_0$\footnote{Time relative to $t_{\rm peak}$, the time when $\mathcal{L}(t)$ peaks, normalized to $t_0=c/(\kappa\rho_0\vt_0^2)$.}&$\mathcal{L}/\mathcal{L}_0\footnote{Energy flux emitted from the surface, normalized to $\mathcal{L}_0=\rho_0\vt_0^3$ (For a surface of $4\pi R^2$ this is equivalent to the luminosity normalized to $4\pi R^2\rho_0\vt_0^3$).}~~(n=3)\footnote{Density is assumed to increase away from the surface as a power law $\rho\propto x^{n}$.}$&$\mathcal{L}(n=1)/\mathcal{L}(n=3)$&$\mathcal{L}(n=3/2)/\mathcal{L}(n=3)$&$\mathcal{L}(n=10)/\mathcal{L}(n=3)$\\
\hline
 -2.00 & 0.00182 & 1.45 & 1.26 & 1.26 \\
 -1.75 & 0.00712 & 1.14 & 1.08 & 1.34 \\
 -1.50 & 0.0236 & 0.979 & 0.984 & 1.34 \\
 -1.25 & 0.0665 & 0.908 & 0.941 & 1.28 \\
 -1.00 & 0.157 & 0.903 & 0.939 & 1.17 \\
 -0.75 & 0.31 & 0.949 & 0.968 & 1.06 \\
 -0.50 & 0.5 & 1.02 & 1.01 & 0.962 \\
 -0.25 & 0.66 & 1.1 & 1.05 & 0.902 \\
  0.00 & 0.719 & 1.13 & 1.07 & 0.882 \\
  0.25 & 0.669 & 1.1 & 1.06 & 0.899 \\
  0.50 & 0.557 & 1.04 & 1.02 & 0.939 \\
  0.75 & 0.435 & 0.99 & 0.993 & 0.989 \\
  1.00 & 0.333 & 0.957 & 0.974 & 1.03 \\
  1.25 & 0.257 & 0.946 & 0.967 & 1.06 \\
  1.50 & 0.201 & 0.948 & 0.969 & 1.07 \\
  1.75 & 0.162 & 0.958 & 0.977 & 1.07 \\
  2.00 & 0.133 & 0.97 & 0.986 & 1.06 \\
  2.25 & 0.112 & 0.983 & 0.997 & 1.05 \\
  2.50 & 0.0963 & 0.996 & 1.01 & 1.04 \\
  2.75 & 0.0839 & 1.01 & 1.02 & 1.02 \\
  3.00 & 0.074 & 1.02 & 1.02 & 1.01 \\
  3.25 & 0.0661 & 1.03 & 1.03 & 1.0 \\
  3.50 & 0.0595 & 1.04 & 1.04 & 0.99 \\
  3.75 & 0.0541 & 1.04 & 1.04 & 0.981 \\
  4.00 & 0.0494 & 1.05 & 1.05 & 0.973 \\
  4.25 & 0.0455 & 1.06 & 1.05 & 0.966 \\
  4.50 & 0.042 & 1.06 & 1.06 & 0.96 \\
  4.75 & 0.039 & 1.07 & 1.06 & 0.954 \\
  5.00 & 0.0364 & 1.08 & 1.06 & 0.95 \\
  6.00 & 0.0285 & 1.1 & 1.08 & 0.934 \\
  8.00 & 0.0194 & 1.14 & 1.09 & 0.913 \\
 10.00 & 0.0144 & 1.17 & 1.11 & 0.9 \\
 15.00 & 0.00845 & 1.22 & 1.13 & 0.879 \\
 20.00 & 0.00578 & 1.26 & 1.14 & 0.867 \\
 30.00 & 0.00338 & 1.3 & 1.16 & 0.851 \\
 50.00 & 0.00172 & 1.33 & 1.18 & 0.836 \\
 75.00 & 0.001 & 1.36 & 1.2 & 0.825 \\
100.00 & 0.000687 & 1.38 & 1.21 & 0.817 \\
125.00 & 0.000513 & 1.41 & 1.22 & 0.811 \\
\hline
\end{tabular}
\caption{Emitted flux\label{tab:L_of_t}}
\end{table*}

\bibliographystyle{apj}

\end{document}